\documentclass[aps,twocolumn,superscriptaddress]{revtex4-2}   
\usepackage{graphicx,amssymb}
\usepackage{color,ulem}
\usepackage{bm}   

\usepackage{bm}

\usepackage {amsmath}
\usepackage{epstopdf}
\usepackage{graphicx,amssymb}
\usepackage{color,ulem}
\usepackage{float}
\usepackage{siunitx}
\usepackage{mathrsfs} 
\usepackage{soul}
 
\begin{document}

\title{Ultrasensitive calorimetric detection of single photons from qubit decay} 
\author{Jukka P. Pekola}
\affiliation{Pico group, QTF Centre of Excellence, Department of Applied Physics, Aalto University School of Science, P.O. Box 13500, 00076 Aalto, Finland}

\affiliation{Moscow Institute of Physics and Technology, 141700 Dolgoprudny, Russia}

\author{Bayan Karimi}
\affiliation{Pico group, QTF Centre of Excellence, Department of Applied Physics, Aalto University School of Science, P.O. Box 13500, 00076 Aalto, Finland}

\affiliation{QTF Centre of Excellence, Department of Physics,
Faculty of Science, University of Helsinki, FI-00014 Helsinki, Finland}
\date{\today}

\begin{abstract}
We describe a qubit linearly coupled to a heat bath, either directly or via a cavity. The main focus of the paper is on calorimetric detection in a realistic circuit, specifically a solid-state qubit coupled to a resistor as an absorber. The bath in the model is formed of oscillators initially in the ground state with a distribution of energies and coupling strengths. A direct numerical solution of the Schr\"odinger equation for the full system including up to $10^6$ oscillators in the bath verifies the expected decay process. We address quantitatively the question of separation of the qubit and bath by adding a cavity in between which by detuning allows one to adjust the decay rate into a convenient regime for detection purposes. Most importantly, we propose splitting a quantum to two uncoupled baths and performing a cross-correlation measurement of their temperatures. This technique enhances significantly the signal-to-noise ratio of the calorimeter.
\end{abstract}
\maketitle
\section{Introduction}
Quantum decay, a century-old problem~\cite{Weisskopf,Khalfin,Asher,Facchi,Peshkin,Fonda,spin-bath,charis,Weiss,campo,babu,Brecht}, is experiencing renaissance thanks to advances and increased interest in quantum technology. Atomic and subatomic physics and later materials physics were the main fields of application of quantum theory in the early twentieth century. Towards the end of the millennium, solid-state artificial quantum systems, in form of nanostructures, have turned quantum science largely into quantum engineering~\cite{Will}.
\begin{figure}
	\centering
	\includegraphics [width=\columnwidth] {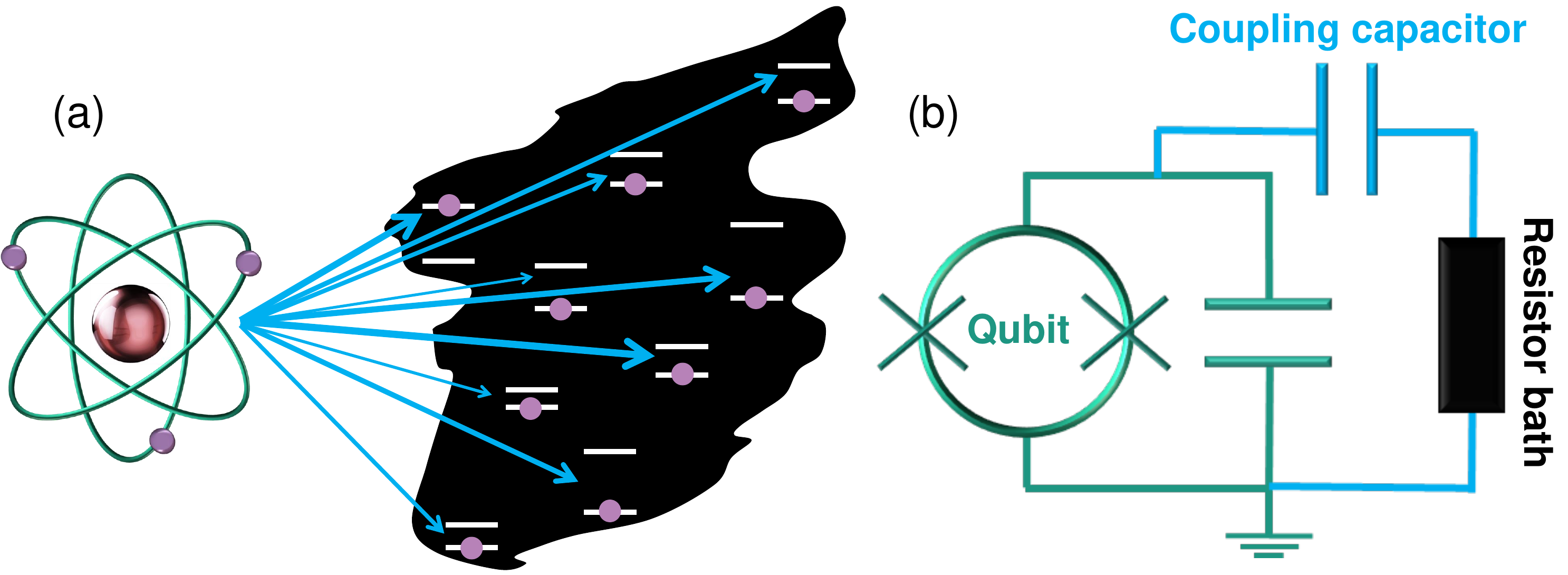}
	\caption{Schematic illustration of an open quantum system. (a) The system is linearly coupled to oscillators forming the bath with distributed couplings and energies. (b) Physical realization of (a) with the corresponding colours, in form of a superconducting qubit coupled capacitively to a resistor. 
		\label{fig1}}
\end{figure}
In this context new aspects of open quantum systems~\cite{Breuer} become interesting, for instance in the domains of quantum thermodynamics and single quantum detection, where decay of open quantum systems is associated with transfer of energy in form of heat~\cite{Deffner,JP,Benenti,Parrondo,Erik,JB}, which we propose here to be detected calorimetrically~\cite{Roukes,xray}. One of the most challenging problems in this context is the continuous detection of low energy photons, emitted by an artificial quantum system, e.g. a superconducting qubit. In this paper we model and analyze the decay of such a system into a heat bath formed of bosonic oscillators. We apply the results on concrete systems, namely on-chip quantum devices coupled to resistive elements as baths. Besides the direct coupling, we analyze a realistic setup where a superconducting cavity is placed between the qubit and the bath. Because of the presence of the cavity, the decay rate can be tuned and it exhibits so-called global and local regimes~\cite{Ronzani}, with a crossover between them determined by the coupling parameters.

As to modelling, it is well recognized that it is not possible to solve the Schr\"odinger equation of the whole “universe”, meaning the system coupled to an infinite reservoir. Instead, one needs to resort to approximate solutions, like weak coupling master equations that find the partial density matrix of the quantum system only, tracing out the environment. This limitation is of course true for an infinitely large reservoir, but one can solve the time evolution of the system and environment exactly for a large but finite number of degrees of freedom in the bath as we do in the current paper. 
By doing this we manage to propose new insight and methods to the currently active line of research of quantum detection, in particular continuous single microwave quantum calorimetry, which presents a still elusive holy grail in experimental research~\cite{Roukes,xray,NETJB,Roope1,Lee}. Several methods have been proposed and tested experimentally and they promise reasonable signal-to-noise ratios in microwave photon detection under ideal conditions~\cite{NETJB,Roope1,Lee,JBPRL,ZBABJ, Walsh,JukkaNJP,Efetov, Oleg,Nakamura,Lescanne,Simone,JBPRL,Govenius}. Here we demonstrate a new approach based on splitting the photon energy to two uncorrelated baths and measuring the coincident absorption by a cross-correlation technique. We demonstrate theoretically that the signal-to-noise ratio of detecting a photon can be enhanced significantly, meaning that even with less ideal measuring setup, as often is the case in a real experiment, one can still resolve these events and the energy of the absorbed photon.  

In Section~\ref{section2} we start by presenting the model in a general setting. Description of technicalities of solving directly the Schr\"odinger equation numerically follows next together with solutions. In Section~\ref{section2C} we discuss the importance of randomness of the parameters for the reservoir to act as a true heat bath. Section~\ref{section2} serves, among other things, as a sanity check of the model we use. Section~\ref{section3} deals with physical realizations. In Section~\ref{section3A} we start by a discussion of electrical circuits. Specifically, we make a connection of the “abstract” coupling coefficients and the spectrum of the oscillators to the concrete circuit parameters. The rest of the paper addresses the problem of detecting small packets of energy of a quantum system, i.e. single photon detection by calorimetry. Some aspects of this question have been theoretically addressed, e.g. in our recent work with a realistic model for the calorimeter~\cite{JBPRL}. In order to consider the actual experimental setup, we place a harmonic cavity between the quantum system and the reservoir in Section~\ref{section3C}. In such a setup based on the hierarchy of coupling strengths in the system, in particular the internal couplings among the quantum elements and those to the heat bath, one can observe the full cross-over between different regimes and address the decay dynamics (Section~\ref{Global and local regimes}). One can also adjust the decay rate at will by Purcell-like detuning of the cavity and qubit and change the decoherence rate in all regimes. Last but not least, we propose decay of the qubit into multiple baths in order to boost the detection efficiency in a cross-correlation measurement. This is a somewhat enigmatic issue of whether a single quantum released by a qubit can be observed simultaneously by two detectors operating by the principle of calorimetry (Section~\ref{section3B}). We propose a cross-correlation experiment on two uncorrelated thermal absorbers in Section~\ref{Cross-correlation of temperatures}. In addressing these problems, we mimic a real heat bath by including a large number (up to $10^6$) of oscillators, bringing it closer to a resistor bath with $\sim 10^8$ degrees of freedom (electrons) in a typical experiment~\cite{NETJB}. The main benefit and motivation of using a large number of oscillators is, however, that we avoid unphysical population revivals and the results become statistically stable. The paper is closed in Section~\ref{section4} by a summary and a list of open questions.

\section{Describing the open quantum system}\label{section2}
\subsection{The model} \label{section2a}
We first consider a qubit with level separation $\hbar\Omega$ coupled to a bath of $N$ oscillators with energy of the $i$:th oscillator equal to $\hbar\omega_i$.
The Hamiltonian reads then
\begin{equation} \label{Hamilton}
\mathcal{\hat{H}} = \hbar\Omega \hat{a}^\dagger\hat{a}+\sum_{i=1}^N \hbar\omega_i \hat{b}_i^\dagger \hat{b}_i+\sum_{i=1}^{N}\gamma_i(\hat{a}^\dagger\hat{b}_i+\hat{a}\hat{b}_i^\dagger),
\end{equation}
where $\hat{a}=|g\rangle \langle e|$ for the qubit with eigenstates $|g\rangle$ (ground) and $|e\rangle$ (excited) and $\hat{b}_i^\dagger~(\hat{b}_i)$ is the creation (annihilation) operator of the oscillator $i$ in the bath. The first two terms form the non-interacting Hamiltonian $\mathcal{\hat{H}}_0=\hbar\Omega \hat{a}^\dagger\hat{a}+\sum_{i=1}^N \hbar\omega_i \hat{b}_i^\dagger \hat{b}_i$ of the qubit and oscillator bath, respectively. The third term represents the coupling between the qubit and the oscillators in the bath as $\hat{V}=\sum_{i=1}^{N}\gamma_i(\hat{a}^\dagger\hat{b}_i+\hat{a}\hat{b}_i^\dagger)$. We take positive real and random valued $\gamma_i$ with uniform distribution from 0 to their maximal value, $\gamma_{i,{\rm max}}$, unless otherwise stated. They incorporate the coupling of the qubit with each oscillator in the bath, shown by the blue arrows in Fig.~\ref{fig1}(a). Later in this section we discuss the importance of randomness in environment parameters from decoherence point of view. For the concrete examples in what follows we typically take a flat distribution of oscillator energies around $\Omega$. The perturbation $\hat{V}$ in the interaction picture with respect to $\mathcal{\hat{H}}_0$ reads 
\begin{eqnarray} \label{Vinter}
\hat{V}_I(t)=\sum_{i=1}^{N}\gamma_i(\hat{a}^\dagger\hat{b}_ie^{i(\Omega-\omega_i)t}+\hat{a}\hat{b}_i^\dagger e^{-i(\Omega-\omega_i)t}).
\end{eqnarray}
We aim to solve the Schr\"odinger equation $i\hbar \partial_t|\psi_I(t)\rangle=\hat{V}_I(t)|\psi_I(t)\rangle$ of the whole system including the bath oscillators. The basis of the Fock states that we use is formed of the states $\{|0\rangle=|1000...0\rangle,~|1\rangle=|0100...0\rangle,...,|i\rangle=|0~0...1^{(\rm i:th)}...0\rangle\}$, where the first entrance in each basis vector refers to the qubit and from the second on to each of the $N$ oscillators in the bath. We thus exclude multiple occupations. The initial state of the whole system (qubit and oscillator bath) is $|\psi_I(0)\rangle\equiv |0\rangle$ meaning that only the qubit is in the excited state $|e\rangle$. This corresponds to the ground state of the bath oscillators but with the qubit excited, e.g., by a $\pi$-pulse.
 
{\bf Justification of the initial state:} The argument in support of the initial ground state approximation lies in the separation of energy scales in the system under study. This justification holds for experiments on nanocalorimetry, where lower $T$ yields improved operation~\cite{NETJB}. For superconducting qubits we have typically $\hbar\Omega/k_{\rm B} \sim 0.3 ... 1$ K, whereas the temperature in an experiment, $T\sim 0.01$ K. This makes it very unlikely to excite relevant environmental modes, since $e^{-\hbar\Omega/k_{\rm B}T} \lesssim 10^{-13}$. Very low energy oscillators, if any, can be thermally excited, but they couple to the qubit by a factor $\sim (\gamma_i/\hbar\Omega)^2$ more weakly than the resonant ones, and can thus be considered as isolated, since in most of our realistic simulations $\gamma_i/\hbar\Omega < 10^{-4}$. This narrows down the Hilbert space in the treatment to a single microstate coupled to states with no or maximally one excitation. By doing this, we can model the whole system. All this is possible only because of the particular but experimentally fully relevant situation where the bath is essentially at zero temperature as compared to the energy of the qubit. In contrast, at temperatures comparable to the qubit energy finding an appropriate basis would seem to be a formidable task because of the astronomically large number of possible microstates of the bath.

In the given basis, the time evolution of the state of the whole system, $|\psi_I(t)\rangle=\sum_{i=0}^{N}\mathscr{C}_i(t)|i\rangle$, follows 
\begin{eqnarray}\label{ampsint}
&&i\hbar\dot{\mathscr{C}}_0(t)=\sum_{j=1}^{N}\gamma_j e^{i(\Omega-\omega_j)t}\mathscr{C}_j(t)\\&&i\hbar\dot{\mathscr{C}}_i(t)=\gamma_i e^{-i(\Omega-\omega_i)t}\mathscr{C}_0(t).\nonumber
\end{eqnarray}
Thus, for instance, $\mathscr{C}_0(t)$ is the amplitude and $|\mathscr{C}_0(t)|^2$ the population of the qubit, and similarly for the bath oscillators with $i=1,2,...,N$. In what follows, when we discuss numerical results we refer to direct integration of Eqs.~\eqref{ampsint} for the present system and later Eqs.~\eqref{ampsint2} for the hybrid one. We find the following integro-differential equation governing the amplitudes $\mathscr{C}_i(t)$ 
\begin{eqnarray}\label{cit}
&&i\hbar\mathscr{C}_i(t)=\int_{0}^{t}dt' \gamma_{i} e^{-i(\Omega-\omega_i)t'}\\&&+\int_{0}^{t}dt' \gamma_{i} e^{-i(\Omega-\omega_i)t'}\int_{0}^{t'}dt''\sum_{j=1}^{N}\gamma_{j}e^{i(\Omega-\omega_j)t''}\mathscr{C}_j(t''),\nonumber
\end{eqnarray} 
with $\mathscr{C}_0(0)=1$ and $\mathscr{C}_i(0)=0$ for $i=1,2,...,N$, consistent with our initial condition. It is worth making a note that we do not impose an artificial boundary between a quantum system and classical environment (“Heisenberg cut”). Instead the whole entity is described by a wave function but thanks to the randomness of the environment  parameters, the oscillators form a heat bath. The decay theory connects in a straightforward manner to the main theme of this paper, the thermal detection of single quanta. Namely, the energy transferred from the qubit to the bath (oscillators) at time $t$ reads $\delta E (t) =\sum_i \hbar \omega_i |\hat b_i(t)|^2$, which by energy conservation equals $\hbar\Omega(1-|\mathscr{C}_0(t)|^2)$.  

We have realized a direct numerical solution of the Schr\"odinger equation for the full system including up to $10^6$ oscillators in the bath and analytic solutions of it. After the initial excitation of the qubit, the decay process in different time intervals verifies quantum decay in short time quadratic (Zeno), long time exponential and eventually power law relaxation regimes~\cite{	Facchi,Peshkin,Fonda,campo}. An example of such a simulation is shown by the blue line in Fig.~\ref{fig-decays} based on the given parameters. For comparison, one can solve Eq.~\eqref{cit} iteratively, which corresponds to perturbation analysis with weak coupling of $\gamma_i$. Based on the mentioned initial conditions, the population of the qubit in the lowest order $|\mathscr{C}_0^{(0)}(t)|^2\equiv 1-\sum_{i=1}^{N}|\mathscr{C}_i^{(0)}(t)|^2$ for the two different regimes is as follows. For the short times, $\Delta\omega t\ll 1$, where $\Delta\omega$ denotes the width of the uniform distribution of $\omega_i$ symmetrically around $\Omega$, we obtain $|\mathscr{C}_0^{(0)}(t)|^2=1-\Lambda_0^2~t^2$ with $\Lambda_0^2=\sum_{i=1}^{N}\gamma_i^2/\hbar^2\equiv N\langle \gamma_{i}^2\rangle/\hbar^2$, where $\langle . \rangle$ denotes the average over all the oscillators. This result represents the quadratic “Zeno” result. For longer times, one obtains linear dependence based on the present approximation as
\begin{equation}\label{cit2_3}
|\mathscr{C}_0^{(0)}(t)|^2=1-\Gamma_0 t,
\end{equation}  
where  
\begin{equation}\label{rate}
		\Gamma_0=\frac{2\pi}{\hbar^2}\sum_{i=1}^{N}\gamma_{i}^2\delta(\omega_i-\Omega)=\frac{2\pi}{\hbar^2}\nu_0\langle \gamma_i^2\rangle
\end{equation} 
denotes the actual decay rate of the qubit. Here $\nu_0=N/\Delta\omega$ is the density of oscillators in $\omega\equiv \omega_i -\Omega$. The last step in Eq.~\eqref{rate} applies for a uniform distribution of oscillators as given above. Equation~\eqref{cit2_3} presents the linear approximation of the exponential decay; the latter can be obtained by standard perturbation theory as well. At times $\Gamma_0 t\gg \ln{({\Delta\omega}/{\Gamma_0})}$ the decay turns into power law in time~\cite{Asher,Peshkin,Rothe}.

\subsection{The significance of randomness}\label{section2C}
\begin{figure}
	\centering
	\includegraphics [width=\columnwidth] {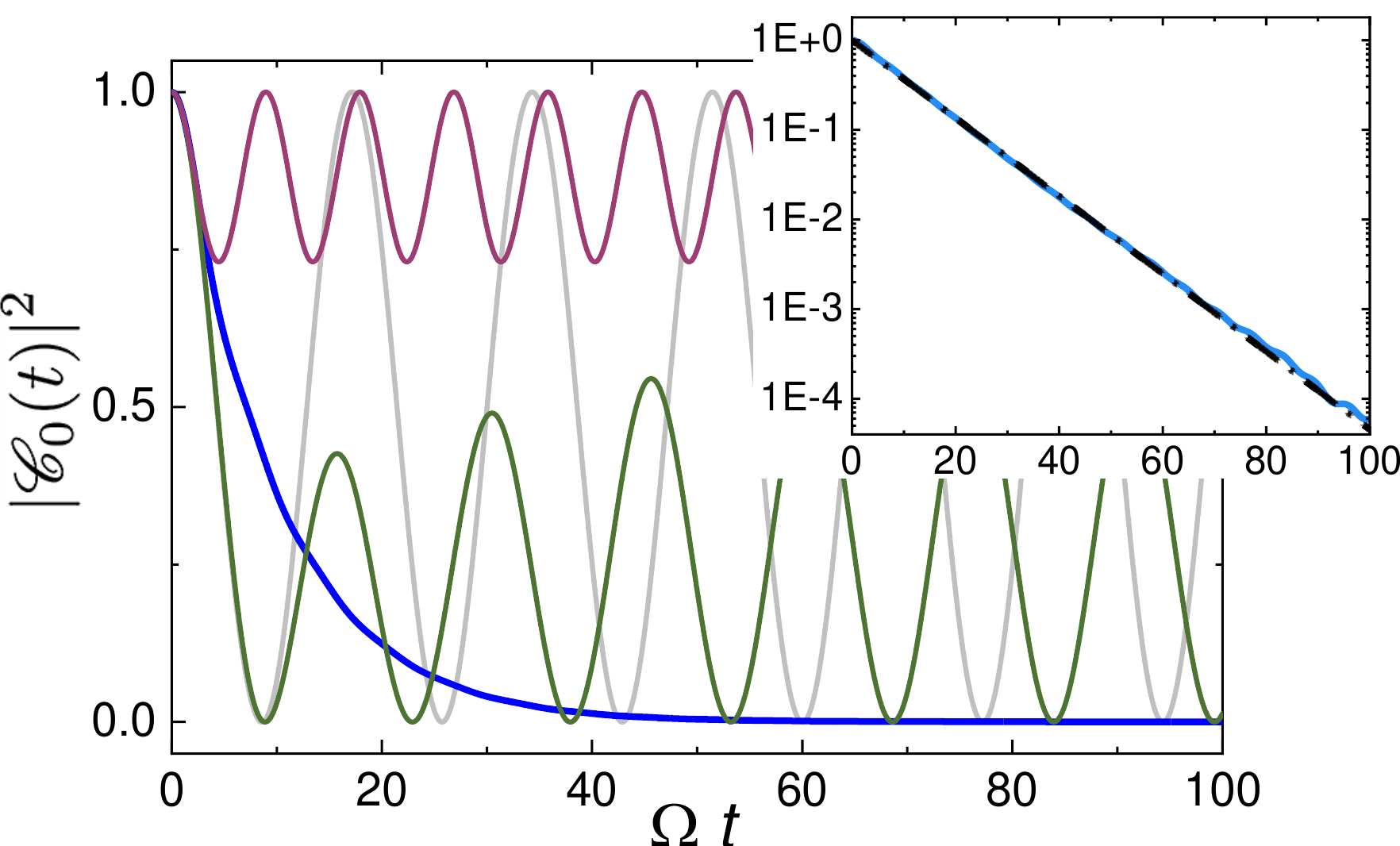}
	\caption{Decay of the qubit population, $|\mathscr{C}_0(t)|^2$, calculated numerically based on the presented model. We assume $N=10^5$ oscillators with uniform distribution of couplings between the qubit and reservoir $0\leq \gamma_{i}\leq \gamma_{i,{\rm max}}$, with overall decay rate $\Gamma_0=0.1\Omega$. The blue decaying line shows the fully random case with uniform distribution of energies of the oscillators ($\Delta\omega=2\Omega$). Inset: The dash dotted line shows the exponential decay, $|\mathscr{C}_0(t)|^2\simeq \exp(-\Gamma_0 t)$ closely following the numerical result shown by solid blue line calculated as in the main panel but with $N=10^6$ for increased accuracy. The rest of the curves in the main panel demonstrate the significance of the randomness of reservoir parameters and couplings on qubit decay. The grey line (pure $\cos^2(\Lambda_0 t)$) is the result for $\Delta\omega=0$, i.e. for oscillators which all have the same energy $\omega_k=\Omega$. The green line presents an intermediate case with $\Delta\omega=0.3\Omega$ with partial revivals. Finally the dark red line is for the oscillators all having the frequency $\omega_k=0.4\Omega$. In this case the population of the qubit never vanishes but oscillates between a non-zero value and unity. All the traces are aligned at short times up to $\Omega t\simeq 3$. 
		\label{fig-decays}} 
\end{figure}

It is in place here to note some properties of the oscillator reservoir and couplings that make it represent a proper heat bath. The general natural principle is randomness. Lack of dispersion in the bath parameters can lead to collective coherent dynamics of the whole circuit with revivals of qubit population. To see that we combine Eqs.~\eqref{ampsint} to obtain an integro-differential equation for $\mathscr{C}_0(t)$
\begin{eqnarray}\label{rebath}
&&\ddot{\mathscr{C}}_0(t)+\Lambda_0^2~\mathscr{C}_0(t)=\\&&-\frac{i}{\hbar^2}\sum_{k=1}^{N}\gamma_{k}^2(\Omega-\omega_k)\int_{0}^{t}dt' e^{i(\Omega-\omega_k)(t-t')}\mathscr{C}_0(t').\nonumber
\end{eqnarray}
Thus, for oscillators with equal level separation $\omega_k\equiv\Omega$ for all, the right-hand side vanishes and the qubit does not decay, even when the couplings $\gamma_i$ are fully random, but it oscillates with population $|\mathscr{C}_0(t)|^2=\cos^2(\Lambda_0t)$. Initial decrease of the qubit population at short times could in such a case be misinterpreted as decay although in reality it precedes the first inevitable revival. Numerics show that imposing even weaker "regularities" in the reservoir leads to non-decaying population of the qubit. This happens for example when the oscillators have equal energies but detuned from the qubit, i.e. $\omega_k\neq \Omega$ (see Fig.~\ref{fig-decays}), or when all the couplings are equal even when oscillators have a distribution of energies (not shown in Fig.~\ref{fig-decays}). 

\section{Physical realizations and circuit quantum thermodynamics}\label{section3}

The central task of this paper is to describe physical systems applicable in experiments on calorimetric quantum detection at low temperatures. For this purpose we present in the next subsection the relation of the coupling coefficients $\gamma_i$ and the spectrum of oscillators to the parameters of a quantum circuit where the heat bath is formed by a resistor~\cite{JBPRL,Matti}. With this background we then move on to address two exciting topics related to experimental setups. 

\subsection{Relation between oscillator properties and circuit parameters}\label{section3A}

The question of how to relate the parameters of practical physical systems to the theoretical ones in the context of open quantum systems is often overlooked in literature. Here we do this for a quantum circuit where a resistor (with resistance $R$) forms the actual bath for a superconducting qubit~\cite{JBPRL,Leggett,Devoret}, see Fig.~\ref{fig1}(b). To start with, we write the phase operator of the oscillator bath in the interaction picture as 
\begin{equation} \label{phase}
\hat \varphi_I (t) =\sum_{i=1}^{N}\lambda_i(\hat{b}_ie^{-i\omega_it}+\hat{b}_i^\dagger e^{i\omega_it}).
\end{equation}
The voltage fluctuations are related to phase as $\delta v(t) = \frac{\hbar}{e}\dot{\hat \varphi} (t)$. The linear coupling is obtained from the voltage noise of the oscillator bath as $\hat{V}(t)=\hat{q}\delta v(t)$ where $\hat{q}=-i\sqrt{\frac{\hbar}{2Z_{\rm Q}}}(\hat{a}-\hat{a}^\dagger)$ is the charge operator of the qubit. Here $Z_{\rm Q}=\sqrt{L_{\rm Q}/C_{\rm Q}}$ where $L_{\rm Q}$ and $C_{\rm Q}$ are the Josephson inductance and capacitance of the qubit, respectively. Then we have in the interaction picture
\begin{equation}\label{voltage-noise}
\hat{V}_I(t)=\sum_{i=1}^{N}\frac{\hbar}{e}\sqrt{\frac{\hbar}{2Z_{\rm Q}}}\omega_i \lambda_i[\hat{a}\hat{b}_i^\dagger e^{-i(\Omega-\omega_i)t}+\hat{a}^\dagger\hat{b}_i e^{i(\Omega-\omega_i)t}].
\end{equation}
Comparing this equation with Eq.~\eqref{Vinter}, we have $\gamma_{i}=\frac{\hbar}{e}\sqrt{\frac{\hbar}{2Z_{\rm Q}}}\omega_i \lambda_i$. In order to relate $\gamma_i$ to $R$ we calculate the spectral density of noise $S_v(\omega) =\int dt e^{i\omega t}\langle \delta v(t)\delta v(0)\rangle$. We identify $\langle\hat{b}_i^\dagger\hat{b}_i\rangle=n(\omega_i)=1/(e^{\beta\hbar\omega_i}-1)$, the Bose distribution at inverse temperature $\beta=(k_{\rm B}T)^{-1}$. $S_v(\omega)$ at $\omega=+\Omega$ is then given by
\begin{eqnarray}\label{s_omega2}
S_v(+\Omega)=2\pi(\frac{\hbar}{e})^2\nu_0\langle\lambda_i^2\rangle\frac{\Omega^2}{1-e^{-\beta\hbar\Omega}},
\end{eqnarray}
On the other hand, the voltage noise of a bare resistor at the same frequency $+\Omega$ reads $S_v(+\Omega)=2R\hbar\Omega/(1-e^{-\beta\hbar\Omega})$~\cite{Lifshitz}. Comparing this spectral density to Eq.~\eqref{s_omega2}, we obtain $\langle\lambda_i^2\rangle= Re^2/(\pi\hbar\nu_0\Omega)$. In the case of the actual circuit of Fig.~\ref{fig1}(b), the expression of $\langle\gamma_i^2\rangle$ in Eq.~\eqref{Vinter} is to be multiplied by $(C_g/C_\Sigma)^2$, where $C_g$ is the coupling capacitance (blue capacitor in Fig.~\ref{fig1}(b)) and $C_\Sigma=C_g+C_{\rm Q}$. Substituting $\lambda_i$ into $\gamma_i$, we have the relation between the coupling $\gamma_i$ and $\nu_0$ and the physical circuit parameters as
\begin{equation}\label{coupling}
\langle \gamma_i^2\rangle=(\frac{C_g}{C_\Sigma})^2\frac{R}{Z_{\rm Q}}\frac{\hbar^2\Omega}{2\pi\nu_0}.
\end{equation}
\subsection{Qubit-cavity-bath setup}\label{section3C}
Here we analyze quantitatively the decay process in a qubit-cavity setup of Fig.~\ref{fig6}. We consider this realistic setup since it is a most common system in the context of, e.g. superconducting circuits, a pioneering one was presented in~\cite{Wallraff} and heat transport experiments on it were done in~\cite{Ronzani} and reviewed in~\cite{JB}. In this section we show that by placing the cavity (coplanar wave resonator) between the qubit and the absorber, we can by tuning the qubit energy with respect to that of the cavity, e.g. by magnetic field, determine almost at will the decay rate of the quantum circuit. This hybrid configuration allows one to perform single-quantum detection under chosen decoherence conditions, and it addresses fundamental questions discussed in Refs.~\cite{JB,Chiara,Hofer,JoachimPRB,Ronzani,Kosloff,Rivas,Hewgill,Marco-C,Smith,Joachim,Jorden,Magazzu}. 

The Hamiltonian of the qubit-cavity-bath setup, schematically presented in Fig.~\ref{fig6}, is given by
\begin{eqnarray}\label{Hamiltonian1}
	\mathcal{\hat{H}}_{\rm QCB}=&&\hbar\Omega \hat{a}^\dagger\hat{a}+ \hbar\Omega_0\hat{c}^\dagger \hat{c}+\sum_{i=1}^N \hbar \omega_i \hat{b}_i^\dagger \hat{b}_i\nonumber\\&&+g(\hat{a}^\dagger\hat{c}+\hat{c}^\dagger\hat{a})+\sum_{i=1}^{N}\gamma_i (\hat{c}^\dagger\hat{b}_i+\hat{c}\hat{b}_i^\dagger),
\end{eqnarray}
where $\hbar\Omega_0$ is the energy difference between the adjacent states of the cavity, with creation/annihilation operators $\hat{c}^\dagger$ and $\hat{c}$, respectively, $g$ indicates the coupling between the qubit and cavity, and $\gamma_i$ for those between cavity and bath oscillators.

We can now isolate the part of the Hamiltonian describing the couplings as $\hat{V}_{\rm QCB}=g(\hat{a}^\dagger\hat{c}+\hat{c}^\dagger\hat{a})+\sum_{i=1}^{N}\gamma_i (\hat{c}^\dagger\hat{b}_i+\hat{c}\hat{b}_i^\dagger)$. Then in the interaction picture corresponding to the non-interacting Hamiltonian $\mathcal{\hat{H}}_{0,\rm QCB}=\hbar\Omega \hat{a}^\dagger\hat{a}+ \hbar\Omega_0\hat{c}^\dagger \hat{c}+\sum_{i=1}^N \hbar \omega_i \hat{b}_i^\dagger \hat{b}_i$ we have the solution $|\psi_I\rangle=(\mathscr{C}_{\rm Q}(t)~\mathscr{C}_{\rm C}(t)~\mathscr{C}_1(t)~\mathscr{C}_2(t)~...)^{\rm T}$ in the basis $\{|1~0~0~0...\rangle,~~|0~1~0~0...\rangle,|0~0~1~0...\rangle,...\}$. Here the left-most index refers to occupation of the qubit, next one to the cavity, and the rest to environmental oscillators. With these conventions we find 
\begin{eqnarray}\label{ampsint2}
	&&i\hbar\dot{\mathscr{C}}_{\rm Q}(t)=g e^{i(\Omega-\Omega_0)t}\mathscr{C}_{\rm C}(t)\nonumber\\&&i\hbar\dot{\mathscr{C}_{\rm C}}(t)=g e^{-i(\Omega-\Omega_0)t}\mathscr{C}_{\rm Q}(t)+\sum_{k}\gamma_k e^{i(\Omega_0-\omega_k)t}\mathscr{C}_k(t)\nonumber\\&&i\hbar\dot{\mathscr{C}_j}(t)=\gamma_j e^{-i(\Omega_0-\omega_j)t}\mathscr{C}_{\rm C}(t).
\end{eqnarray}
Again these equations are solved numerically for a similar oscillator bath as in Section~\ref{section2}, yielding the general solution in the low temperature limit.

\subsection{Decay in the global and local regimes}\label{Global and local regimes} 
\begin{figure*}
	\centering
	\includegraphics [width=\textwidth] {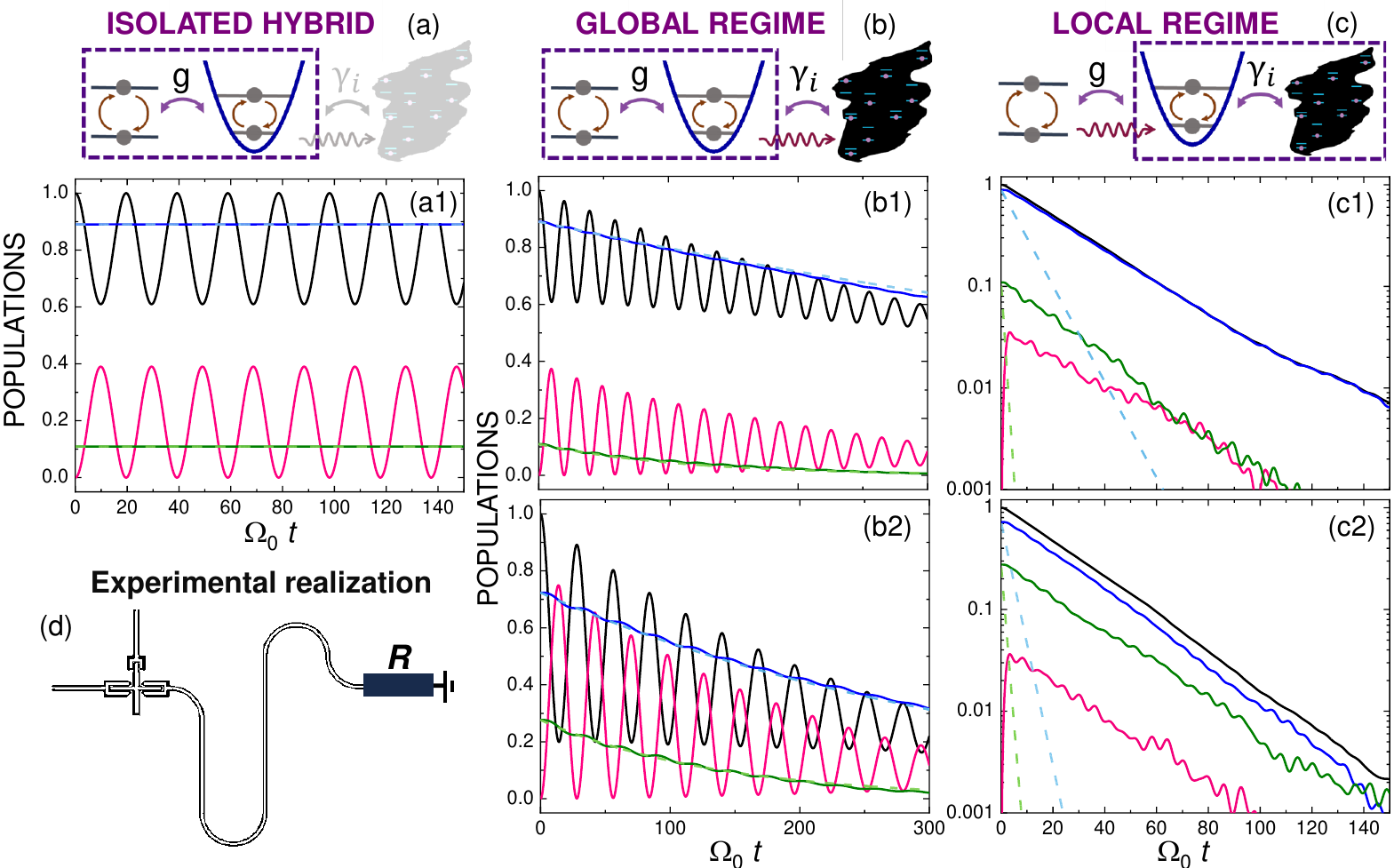}
	\caption{Different relaxation regimes of a qubit coupled via a cavity to a heat bath. The qubit (alone) has been initially prepared in its excited state. Here the coupling between the qubit and cavity is $\bar{g}=0.1$, and the number of oscillators in the bath is $N=10^4$ with flat energy spectrum of width $\Delta \omega=2\Omega_0$. In the three panels (a1),(b1), and (c1) the qubit is detuned at $D=-0.25$ and in panels (b2) and (c2) $D=-0.1$. The black solid line is the population of the qubit, $|\mathscr{C}_{\rm Q}(t)|^2$, the red solid line the population of the cavity, $|\mathscr{C}_{\rm C}(t)|^2$, blue solid line the population in the first eigenstate of the hydrid system, $\rho_{11}$, and the green one that of the second eigenstate, $\rho_{22}$. Finally the light blue and green dashed lines are the predictions of the global model for $\tilde{\rho}_{11}$ and $\tilde{\rho}_{22}$, respectively. In panel (a1) $\Gamma_0 = 0$, i.e. it presents an isolated hybrid system, as shown schematically in panel (a), with populations $\rho_{11}~(=\tilde{\rho}_{11})$ and $\rho_{22}~(=\tilde{\rho}_{22})$ from master equation (numerically from Eqs.~\eqref{pops}). In both panels (b1) and (b2) $\Gamma_0=0.01\Omega_0$ which presents exponential decay of the population in hybridized states as expected for global picture as shown schematically in (b). In panels (c1) and (c2), coupling to the bath is strong, $\Gamma_0=\Omega_0$, which leads to the breakdown of the global picture and entrance into the local regime with exponential decay of the qubit only (local picture, as shown in (c)). (d) A possible physical realization of the system.   
		\label{fig6}}
\end{figure*}

Here we analyze two relevant approximations of the hybrid "qubit-cavity-bath" entity either in a {\it global} or {\it local} picture~\cite{JB,Chiara,Hofer,JoachimPRB,Ronzani,Kosloff,Rivas,Hewgill,Marco-C,Smith,Joachim,Jorden,Magazzu} as schematically shown on top panels of~\ref{fig6}(b) and~\ref{fig6}(c). In the global view, the qubit and cavity form a quantum hybrid which is then weakly coupled to the heat bath, whereas in the local view the qubit only forms the quantum system that decays to the bath via the cavity that gives extra spectral filtering. The validity of the picture in general depends on where the bottleneck of coupling is.

In the global approximation of our hybrid (top panels of~\ref{fig6}(b)), the dimensionless Hamiltonian, normalized by $\hbar\Omega_0$, for the bare system in the absence of environment oscillators is given by
\begin{eqnarray}\label{H0GVG}
	\mathcal{\hat{H}}_{\rm 0,G}=r \hat{a}^\dagger\hat{a}+ \hat{c}^\dagger \hat{c}+\bar{g}(\hat{a}^\dagger\hat{c}+\hat{c}^\dagger\hat{a})
\end{eqnarray}
and the perturbation as
\begin{equation}\label{H0GVG2}
	\hat{V}_{\rm G}=\sum_{i=1}^{N}\bar{\gamma}_i(\hat{c}^\dagger\hat{b}_i+\hat{c}\hat{b}_i^\dagger)
\end{equation}
where $r=\Omega/\Omega_0$, $\bar{g}=g/\hbar\Omega_0$ and $\bar{\gamma}_i={\gamma}_i/\hbar\Omega_0$. We employ the product basis for the hybrid as $\{|00\rangle,$ $|10\rangle,$ $|01\rangle\}$ where the first entry refers to the qubit and the next one to the cavity.

The eigenenergies $\varepsilon_{i}$ of the Hamiltonian~\eqref{H0GVG}, also normalized by $\hbar\Omega_0$ are given by
\begin{eqnarray}\label{energies}
	\varepsilon_0=0,~~\varepsilon_{1,2}=1+\frac{1}{2} [D\mp\sqrt{D^2+4\bar{g}^2}],
\end{eqnarray}
where the $-$ and $+$ signs refer to $\varepsilon_1$ and $\varepsilon_2$, respectively, and $D\equiv r-1$ is the detuning of the qubit-cavity hybrid. The corresponding eigenstates are
\begin{eqnarray} \label{vectors}
	|\tilde{0}\rangle = \left(
	\begin{array}{ccc}
		1 \\ 0 \\0
	\end{array}
	\right),
	~|\tilde{1}\rangle = \left(
	\begin{array}{ccc}
		0 \\ {\alpha}_1 \\{\alpha}_2
	\end{array}
	\right),
	~|\tilde{2}\rangle = \left(
	\begin{array}{ccc}
		0 \\ {\alpha}_3 \\{\alpha}_4
	\end{array}
	\right).
\end{eqnarray} 
Here ${\alpha}_1=(D-\eta)/{\sqrt{4\bar{g}^2+(D-\eta)^2}}$, ${\alpha}_2={2\bar{g}}/{\sqrt{4\bar{g}^2+(D-\eta)^2}}$, ${\alpha}_3=(D+\eta)/{\sqrt{4\bar{g}^2+(D+\eta)^2}}$,  ${\alpha}_4={2\bar{g}}/{\sqrt{4\bar{g}^2+(D+\eta)^2}}$, and $\eta=\sqrt{D^2+4\bar{g}^2}$. 

The transition rates $\Gamma_{i\rightarrow j}$ between the eigenstates are determined by the matrix elements $\langle \tilde{i}|\hat{q}_{\rm c}|\tilde{j}\rangle$ of the charge operator $\hat{q}_{\rm c}$ of the cavity and the voltage noise $S_v(\Omega_0)$ induced by the bath on the cavity at frequency $\Omega_0$ as 
\begin{equation}\label{trans}
	\Gamma_{i\rightarrow j}=\frac{1}{\hbar^2}|\langle \tilde{i}|\hat{q}_{\rm c}|\tilde{j}\rangle|^2S_v(\Omega_0),
\end{equation}
where $\hat{q}_{\rm c}=-i\sqrt{\frac{\hbar}{2Z_0}}(\hat{c}-\hat{c}^\dagger)$ with $Z_0$ the impedance of the cavity. We assume that the bath couples only to the cavity of the hybrid system (qubit and cavity) which is a realistic assumption, e.g. in superconducting circuit architectures where the physical separations are large. Equilibrium bath induces again voltage noise $S_v(\omega)=2R\frac{\hbar\omega}{1-e^{-\beta\hbar\omega}}$. Then the transition rates at zero bath temperature are
\begin{eqnarray}\label{transitionrates}
	\Gamma_{1\rightarrow 0}=\frac{\Omega_0}{Q}{\alpha}_2^2,~~\Gamma_{2\rightarrow 0}=\frac{\Omega_0}{Q}{\alpha}_4^2.
\end{eqnarray}
If we assume that the noise source is directly connected to the cavity, without coupling capacitor, then the quality factor of the cavity is $Q=Z_0/R$. Other rates vanish: $\Gamma_{1\rightarrow 2}=\Gamma_{2\rightarrow 1}=0$ due to selection rule, and $\Gamma_{0\rightarrow 1}=\Gamma_{0\rightarrow 2}=0$ at $T=0$. Referring to Eqs.~\eqref{rate} and \eqref{coupling}, we identify $\Gamma_0\equiv{\Omega_0}/{Q}$ in this setup. Based on this simple decay scheme we find the populations of the eigenstates $\rho_{11}(t)=\rho_{11}(0)\exp(-\Gamma_{1\rightarrow 0}t)$ and $\rho_{22}(t)=\rho_{22}(0)\exp(-\Gamma_{2\rightarrow 0}t)$, where $\rho_{11}(0)={\alpha}_1^2$ and $\rho_{22}(0)={\alpha}_2^2$. Similarly, since $\dot{\rho}_{00}=\Gamma_{1\rightarrow 0}\rho_{11}+\Gamma_{2\rightarrow 0}\rho_{22}$ for the population $\rho_{00}$ of the ground state, we find immediately the decay rate at $t=0$ to the ground state as $\dot{\rho}_{00}(0)=2\Gamma_0{\alpha}_1^2{\alpha}_2^2$, i.e. 
\begin{equation}\label{rho00}
	\dot{\rho}_{00}(0)=\frac{\Gamma_0/2}{1+(\frac{1}{2\bar{g}})^2D^2}.
\end{equation}
It thus obeys Lorentzian dependence on the detuning of the qubit-cavity with effective quality factor $(2\bar{g})^{-1}$.

In order to assess whether the global treatment works we write the estimates of the populations $\tilde{\rho}_{11}(t)$ and $\tilde{\rho}_{22}(t)$ from the numerical solution of the Schr\"odinger equation in the time-dependent eigenstates $|\tilde{0}\rangle=|0~0\rangle$, $|\tilde{1}\rangle={\alpha}_1e^{-i\Omega t}|1~0\rangle+{\alpha}_2e^{-i\Omega_0 t}|0~1\rangle$, and $|\tilde{2}\rangle={\alpha}_3e^{-i\Omega t}|1~0\rangle+{\alpha}_4e^{-i\Omega_0 t}|0~1\rangle$ as
\begin{eqnarray}\label{pops}
	&&\tilde{\rho}_{11}(t)=|\langle \tilde{1}|\psi_I(t)\rangle|^2=|{\alpha}_1e^{-i\Omega t}\mathscr{C}_{\rm Q}(t)+{\alpha}_2e^{-i\Omega_0 t}\mathscr{C}_{\rm C}(t)|^2\nonumber\\&&\tilde{\rho}_{22}(t)=|\langle \tilde{2}|\psi_I(t)\rangle|^2=|{\alpha}_3e^{-i\Omega t}\mathscr{C}_{\rm Q}(t)+{\alpha}_4e^{-i\Omega_0 t}\mathscr{C}_{\rm C}(t)|^2.\nonumber\\
\end{eqnarray}

We now present quantitatively the cross-over starting from isolated hybrid system via an open global one, and finally to the fully incoherent local qubit with increasing coupling $\Gamma_0$ to the bath. We assume that at $t<0$ the system is in equilibrium at zero temperature in the state where all the oscillators (including the qubit and cavity) are in the ground state. The system is then initialized at $t=0$ in the state $|1~0~0...\rangle$, meaning that the qubit is driven to the excited state. In Fig.~\ref{fig6} we present the numerically solved $|\mathscr{C}_{\rm Q}(t)|^2$, i.e. the population in the excited state of the qubit (black line), $|\mathscr{C}_{\rm C}(t)|^2$, the excited state population of the cavity (red line), and $\tilde{\rho}_{11}(t)$ and $\tilde{\rho}_{22}(t)$ with blue and green solid lines, respectively. The corresponding populations $\rho_{11}(t)$ and $\rho_{22}(t)$ from the master equations of hybrid system are shown by light blue and green dashed lines. Moreover, the top (Figs.~\ref{fig6}\,(a1), \ref{fig6}\,(b1), and \ref{fig6}\,(c1)) and bottom (Figs.~\ref{fig6}\,(b2) and \ref{fig6}\,(c2)) panels correspond to two different values of detuning $D=-0.25$ and $D=-0.1$, respectively.

In the isolated qubit-cavity system ($\Gamma_0\equiv 0$) the populations $|\mathscr{C}_{\rm Q}(t)|^2$ and $|\mathscr{C}_{\rm C}(t)|^2$ of the qubit and cavity oscillate out-of-phase in accordance with the solution of \eqref{ampsint2} for $\gamma_i\equiv 0$ as 
\begin{eqnarray}\label{cqgamma0}
	|\mathscr{C}_{\rm Q}(t)|^2&&=1-|\mathscr{C}_{\rm C}(t)|^2=\frac{1}{2}[1+\frac{D^2}{D^2+4\bar{g}^2}\nonumber\\&&+\frac{4\bar{g}^2}{D^2+4\bar{g}^2}\cos(\sqrt{D^2+4\bar{g}^2} t)].
\end{eqnarray}
Thus $|\mathscr{C}_{\rm Q}(t)|^2$ oscillates between $D^2/(D^2+4\bar{g}^2)$ and $1$. On the other hand, the populations in the eigenstates of the hybrid, $\rho_{11}(t)~(=\tilde{\rho}_{11}(t)$ in this case) and $\rho_{22}(t)~(=\tilde{\rho}_{22}(t))$, remain strictly constant, and their values are determined by the coupling $\bar{g}$ and detuning $D$. This is demonstrated in Fig.~\ref{fig6}(a1). In panels \ref{fig6}(b1) and \ref{fig6}(b2), we introduce weak coupling of the cavity to the bath, $\Gamma_0=0.01\Omega_0$, for two values of detuning. In this situation the numerical solution of Eq.~\eqref{ampsint2} shows that the global description given above applies: populations $\tilde{\rho}_{11}(t)$ and $\tilde{\rho}_{22}(t)$ decay exponentially fully overlapping with $\rho_{11}(t)$ and $\rho_{22}(t)$, respectively, shown also in the figure. On the contrary, the populations of qubit $|\mathscr{C}_{\rm Q}(t)|^2$ and cavity $|\mathscr{C}_{\rm C}(t)|^2$ oscillate, but these oscillations are damped over time scale $\sim \Gamma_0^{-1}$. 

Further increasing $\Gamma_0$ well beyond the internal coupling $\bar {g}$ leads to the failure of the global model. Both panels in Figs.~\ref{fig6}(c1) and \ref{fig6}(c2) present the case where $\Gamma_0=\Omega_0$. In this regime, all the coherent behaviour of the qubit-cavity system has vanished, and the qubit alone, $|\mathscr{C}_{\rm Q}(t)|^2$, follows closely $\tilde{\rho}_{11}(t)$, decaying exponentially from the excited state and the cavity remains mainly in the ground state, $|\mathscr{C}_{\rm C}(t)|^2\simeq 0$. Naturally the predictions of the global model for $\rho_{11}(t)$ and $\rho_{22}(t)$ fail in this regime as shown by the dashed lines. 

Indeed both panels in Figs.~\ref{fig6}(c1) and \ref{fig6}(c2) indicate local regime where the Hamiltonian and perturbative terms are respectively given by $\mathcal{\hat{H}}_{\rm 0,L}=\hbar\Omega \hat{a}^\dagger\hat{a}+ \hbar\Omega_0\hat{c}^\dagger \hat{c}+\sum_{i=1}^N \hbar \omega_i \hat{b}_i^\dagger \hat{b}_i+\sum_{i=1}^{N}\gamma_i (\hat{c}^\dagger\hat{b}_i+\hat{c}\hat{b}_i^\dagger)$ and $\hat{V}_{\rm L}=g(\hat{a}^\dagger\hat{c}+\hat{a}\hat{c}^\dagger)$. To model this behaviour in local regime we calulate the persistence amplitude of the qubit in the Schr\"odinger picture $\mathscr{C}^{\rm (S)}_{\rm Q}(t)=\langle 1~0~0~0...|e^{-i\mathcal{\hat{H}}_{\rm QCB}t/\hbar}|1~0~0~0...\rangle$. Solving it to the second order we have
\begin{eqnarray}\label{interest2_8}
	|\mathscr{C}_{\rm Q}(t)|^2= 1-2\pi\frac{g^2}{\hbar^2}\sum_{i}|c_i|^2\delta(\Omega-E_i/\hbar)t,
\end{eqnarray} 
where $E_i$ is the energy and $c_i$ the projection of the i:th eigenstate of the  $\mathcal{\hat{H}}_{\rm 0,L}$ on the cavity state $|0~1~0~0~...\rangle$. First order perturbation theory for non-degenerate states yields $E_i^{(1)}=\hbar\omega_i$ and $c_i^{(1)}=\frac{\gamma_i}{\hbar\omega_i-\hbar\Omega_0}$. In this case $|\mathscr{C}_{\rm Q}(t)|^2$ is given by
\begin{eqnarray}\label{interest2_10}
	|\mathscr{C}_{\rm Q}(t)|^2=1-\frac{\bar{g}^2\Gamma_0}{D^2}t\equiv 1-\Gamma_{\rm L} t,
\end{eqnarray}
which is valid for $|D|\gg\Gamma_0/\Omega_0$ in agreement with numerics. 

Besides the cross-over between the two decay modalities, we have importantly shown in this section that in all regimes the decay rate of the quantum system can be varied by detuning the qubit and the cavity.

\begin{figure}
	\centering
	\includegraphics [width=\columnwidth] {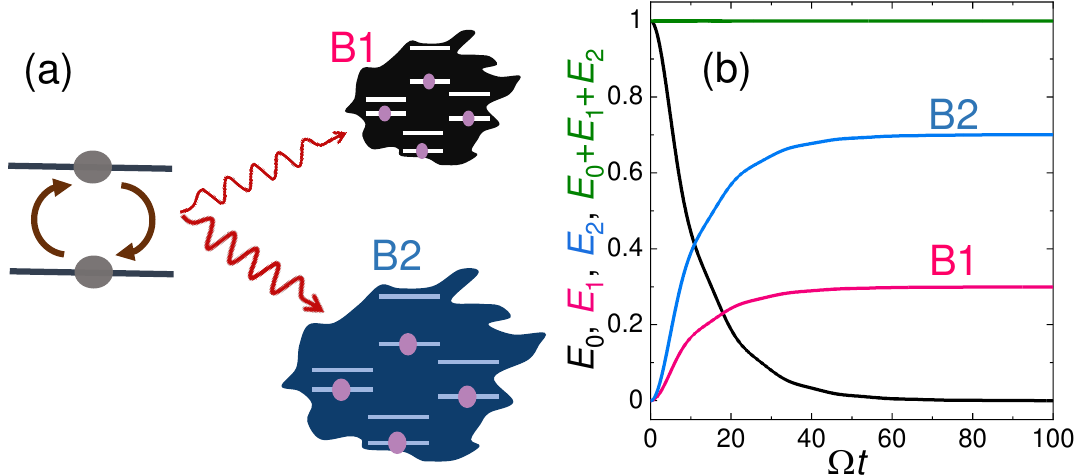}
	\caption{Decay of the qubit to two uncoupled baths. (a) Scheme of the qubit and two baths B1 and B2 presented. (b) Results of solution of Schr\"odinger equation for a qubit coupled to $N=10^6$ bath oscillators of which $N_1=3\times 10^5$ are in B1 and $N_2=7\times 10^5$  in B2. The overall decay rate to the two baths is $\Gamma_0=0.084\Omega$. We use $\Delta\omega=\Omega$ in both baths. The descending line shows the expectation value of energy of the qubit, $E_0=\hbar\Omega|\mathscr{C}_0(t)|^2$, and the two rising lines $E_1=\sum_{i=1}^{N_1}\hbar\omega_i|\mathscr{C}_{i,1}(t)|^2$ and $E_2=\sum_{i=1}^{N_2}\hbar\omega_i|\mathscr{C}_{i,2}(t)|^2$, the expectation values of energy injected to baths B1 and B2. The horizontal line at top demonstrates energy conservation over the whole time of decay, i.e. $E_0+E_1+E_2=\hbar\Omega$.
		\label{fig4}}
\end{figure}

\begin{figure*}
	\centering
	\includegraphics [width=\textwidth] {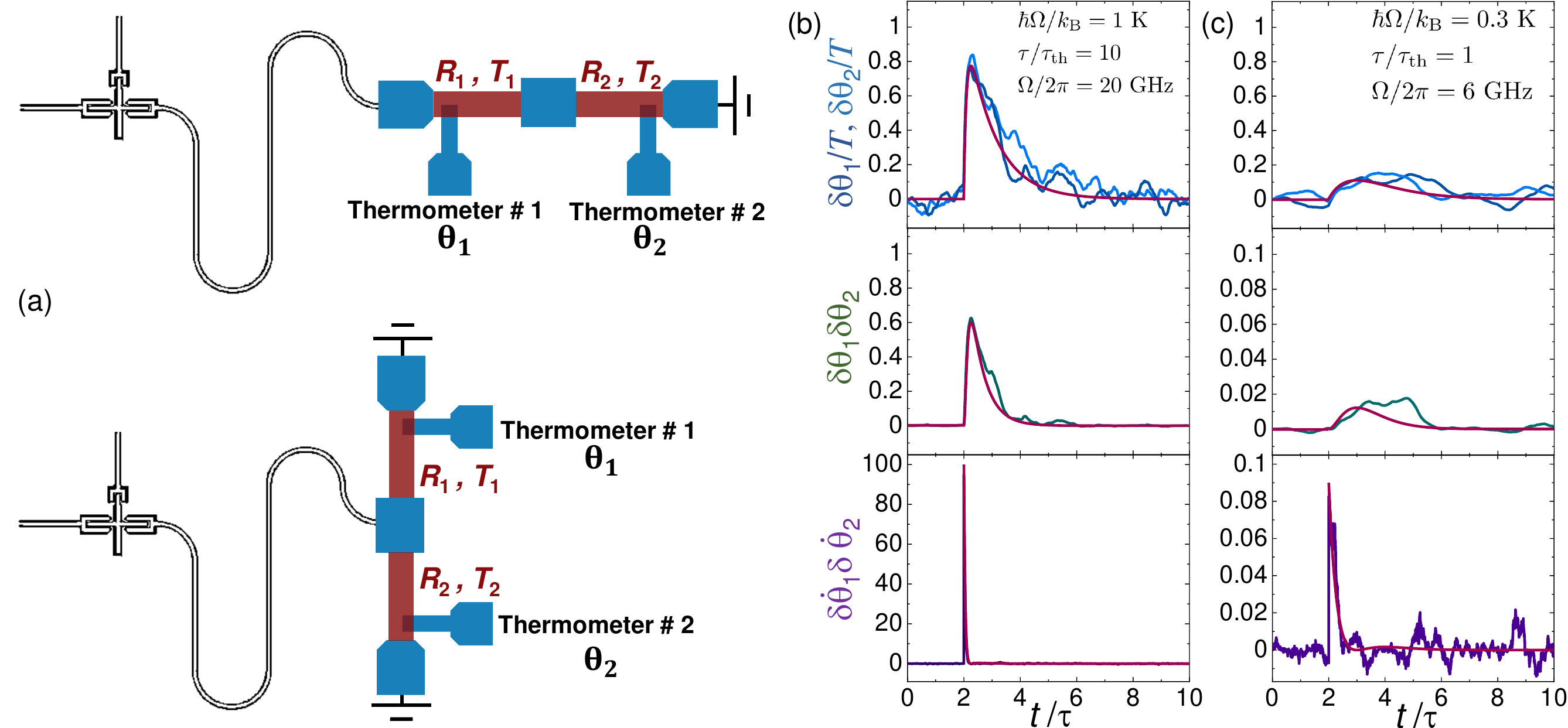}
	\caption{The proposed cross-correlation setup and numerical results on it. (a) A schematic (not to scale) presentation of a transmon type superconducting qubit, on the left, coplanar wave resonator in the middle and the split absorber, on the right. Each half has its own thermometer and they are thermally isolated from each other by a superconducting connection in between. Here blue color refers to the superconductor and brown to normal metal. We present two alternative options of splitting the absorber, a series connection (top) and parallel connection (bottom), with essentially the same characteristics in detection. (b) and (c) Time-dependent temperature of the two thermometers $\theta_i$, in response to 1~K photon with a realistic thermometer ($\tau/\tau_{\rm th}=10$) in (b) and 0.3~K photon with somewhat slower thermometer ($\tau/\tau_{\rm th}=1$) in (c). The results of the two types of correlation measurements, $\theta_1(t)\theta_2(t)$ and especially $\dot\theta_1(t)\dot\theta_2(t)$, demonstrate how the absorption signal can be resolved even in the presence of inevitable thermal noise. The smooth (red) curves in all these panels are analytical results excluding the noise, originating from $\bar{\theta}_i$ in Eq.~\eqref{theta}.
		\label{cross-correlation}}
\end{figure*}

The experimental realization of the qubit-cavity-bath hybrid is presented in Fig.~\ref{fig6}(d), where the system is a transmon qubit coupled in a typical configuration to a coplanar wave resonator as the cavity~\cite{Koch}. One practical way to realize a detector in this setup is to combine the circuit quantum electrodynamics setup with calorimeter formed of an on-chip resistor monitored by a tunnel junction thermometer. In this case the qubit on the left is capacitively $(g)$ coupled to the cavity. The cavity releases the energy to the resistive bath ($R$) over a time given by $Q/\Omega_0$. This is typically of the order of tens or hundreds of nano-seconds for instance in the experiment of Ronzani et al.~\cite{Ronzani}. It is much faster than the relaxation time to the phonon bath, which is typically $10\,-\,100~\mu$s at low temperatures~\cite{NETJB}. Suppose the cavity in Fig.~\ref{fig6}(d) is a $\lambda/4$ resonator with current maximum at its end. Then the terminating resistor on the right dissipates the energy of the resonator in the said time interval $Q/\Omega_0$ which corresponds to $\Gamma_0^{-1}$ in the model above.

\section{Multiple baths}\label{section3B}

Detecting low energy microwave photons calorimetrically one by one, like those emitted by e.g. superconducting qubits, is a challenge. Several ultrasensitive detectors are in principle able to detect quanta in the said microwave regime, but up to now none of them has achieved this result. In this section we propose splitting the energy of a photon to two uncorrelated baths. Both of these baths are equipped with proper thermometers that can monitor the temperature of the corresponding bath continuously. We demonstrate that one can boost significantly the detection efficiency by applying a cross-correlation measurement technique. 

We start by clarifying the splitting concept. Interesting fundamental and practical questions arise when a cavity or qubit is coupled to multiple baths~\cite{Belyansky,Ronzani}. Is the energy of this quantum given to one or can it be split to several baths, and in what way? To settle this question we perform the same analysis as above but now we assume that the $N$ bath oscillators are distributed such that $N_1$ of them form bath B1 and $N_2=N-N_1$ bath B2, as schematically shown in Fig.~\ref{fig4}(a). We assume that there is no direct mutual coupling between the two baths B1 and B2. The time evolution of these oscillators is again determined by Eqs.~\eqref{ampsint}, and the energy given to each bath can be then analyzed accordingly. In particular, for baths that are internally uncoupled we can straightforwardly determine the expectation value of energy of each of them as $E_1=\sum_{k=1}^{N_1}\hbar\omega_k|\mathscr{C}_k(t)|^2$ and $E_2=\sum_{k=N_1+1}^{N}\hbar\omega_k|\mathscr{C}_k(t)|^2$ at time $t$. It follows that for a qubit similarly coupled to each bath (same distribution of $\gamma_{i}$) and with similar distribution of oscillator energies, we expect $E_1$ and $E_2$ to be distributed in proportion to the number of oscillators in each of them. Figure \ref{fig4}(b) presents a numerical example of what is written above with given parameters. 

According to our argument and analysis the energy of the quantum can thus be distributed to multiple baths. This would allow potentially for a significant boost in detection efficiency, e.g. in a calorimetric detection if one measures the temperature of each bath of Fig.~\ref{fig4}(a) simultaneously with a proper thermometer~\cite{NETJB}. By doing a cross-correlation measurement of the two temperatures, one would then enhance the signal-to-noise ratio~\cite{crossnoise}. We may give a simple argument on how the temperature measurement of the two baths can be related to the energies $E_1$ and $E_2$. Let us assume that each thermometer measures the temperature of the corresponding bath at a repetition rate that is faster than the energy release rate to the phonon bath but slower than the energy release of the qubit to the corresponding absorber. We argued in the previous section that such a time-window exists. The combined system, qubit and all the bath oscillators, evolves according to the Schr\"odinger equation between the measurements. In a projective measurement each thermometer reads a temperature $T_i$ that is directly given by $E_i$ via the heat capacity $\mathcal{C}_i$. This yields an abrupt jump in $T_i$ once the photon is absorbed. The decay of the temperatures back to the phonon bath is then essentially a classical process that the thermometers record as will be explained below. 

\subsection{Cross-correlation of temperatures~\textemdash~measurement scheme}\label{Cross-correlation of temperatures}

To make our argument concrete and to demonstrate its experimental feasibility, we apply the temperature cross-correlation method to a circuit presented in Fig.~\ref{cross-correlation}(a). With the two constructions depicted, either series or parallel, having a superconducting connector in between them, the resistors are thermally isolated from each other, and what follows is that their temperature noises are uncorrelated. The two resistors act as “twin absorbers” of the calorimeter with resistances $R_1$ and $R_2$. The energy stored in the resonator and released to the absorbers rapidly, over the time $Q/\Omega_0$, is in this cross-correlation setup distributed among the two absorbers according to elementary circuit principles. In the series configuration, shown in the top panel of Fig.~\ref{cross-correlation}(a), the ratio of energies released to the resistors 1 and 2 equals $R_1/R_2$, whereas in the parallel connection of Fig.~\ref{cross-correlation}(a) bottom, it is $R_2/R_1$. This circuit analysis then gives a way to interpret the outcome of the cross-correlation measurement and the “splitting of the quantum” in accordance with the quantum picture above.

The case of a single absorber was studied in Ref.~\cite{JBPRL}, where the response of a calorimeter to a single 20~GHz (1\,K) photon absorbed instantaneously was analyzed in the presence of inevitable heat current noise due to coupling of the absorber to the phonon bath at $T=0.01$~K. The approach is to use the Langevin equation as 
\begin{equation}\label{Langevin-equation}
\delta\dot{T}_i(t)=-\tau^{-1}\delta T_i(t)+\delta \dot{Q}_i(t)/\mathcal{C}_i
\end{equation}
to produce numerically the two temperature traces with uncorrelated noises for the symmetric case $R_1=R_2$. Here $\delta T_i(t)$ is the deviation of the absorber $i$ temperature from that of the phonon bath, $\tau$ denotes the electron-phonon relaxation time which is about $100\,\mu$s at the lowest temperatures, and $\dot{Q}_i(t)$ is the fluctuating heat current obeying the fluctuation dissipation theorem. In Fig.~\ref{cross-correlation}(b) and \ref{cross-correlation}(c) we present the advantage of using the cross-correlation measurement in this configuration. The top panels show the response (measured relative temperature deviation from that of the bath, $\theta_i(t)$) of the two thermometers separately for the absorption of a quantum with 1~K (Fig.~\ref{cross-correlation}(b)) and 0.3~K (Fig.~\ref{cross-correlation}(c)) energy. The $\theta_i(t)$ is obtained via relaxation time ($\tau_{\rm th}$) approximation from the actual temperatures $T_i$ by solving 
\begin{equation} 
\dot{\theta}_i(t)=-\tau_{\rm th}^{-1}[\theta_i(t)-\delta T_i(t)].
\end{equation} 
Here $\tau_{\rm th}$ is the thermometer response time. Apart from noise, the measured temperature $\theta_i(t)$ follows the expression
\begin{equation}\label{theta}
	\bar{\theta}_i(t)=\Delta T_i\frac{\tau}{\tau-\tau_{\rm th}}(e^{-t/\tau}-e^{-t/\tau_{\rm th}}),
\end{equation}
where $\Delta T_i=\hbar\Omega/\mathcal{C}_i$. In Fig.~\ref{cross-correlation}(b), the single thermometer with response time $\tau_{\rm th}=0.1\tau$, seems to be sufficient for the task of resolving the transient due to the absorption event of a 1~K photon with reasonable signal to noise ratio, in the absence of instrumental excess noise. However, by applying the cross-correlation method, either by taking the product $\theta_1(t)\theta_2(t)$ or especially the product of the derivatives $\dot\theta_1(t)\dot\theta_2(t)$ improves the signal to noise ratio significantly as claimed above. In \ref{cross-correlation}(c) the lower energy of the photon and slower response time $\tau_{\rm th}=\tau$ of the thermometer makes it next to impossible to resolve the 0.3~K photon by a single thermometer. Yet the cross-correlation technique, in particular the $\dot\theta_1(t)\dot\theta_2(t)$ method, would allow one to detect the photon with good signal to noise ratio under these conditions as well.

\section{Summary and outlook}\label{section4}

In this paper we have put the general framework of quantum decay into the context of quantum calorimetry. We first presented the methods used and revisited the common problem of decay of a quantum two-level system by directly solving the Schr\"odinger equation for up to $10^6$ bath oscillators. The heart of the paper deals with the connection of the general picture to real physical systems, and discussion of the properties of a heat bath. Finally we assess the measurement strategies in observing single emission events and the issue of decay of a hybridized quantum system. 

Several tasks remain for future studies. First, the case of finite temperature bath is a somewhat challenging problem as described in the paper. In our present context this is of less importance since the typical energy of a qubit clearly exceeds the thermal energy in a superconducting circuit at millikelvin temperatures. On the practical level, the precise cross-correlation measurement configuration and the projected enhancement of signal-to-noise ratio need to be analyzed in specific setups case-by-case. The correspondence of the model system and the physical one needs naturally specific analysis for a chosen circuit in terms of the type of qubit, and the precise way of coupling (for instance inductive instead of capacitive coupling) of it to the cavity and to the environment. Yet the results obtained here are quite general concerning the response of the calorimeter.

\section{Acknowledgments}
We thank Paolo Muratore-Ginanneschi, Brecht Donvil, Dmitry Golubev and George Thomas for useful discussions. This work was funded through Academy of Finland grant 312057 and from the European Union's Horizon 2020 research and innovation programme under the European Research Council (ERC) programme and Marie Sklodowska-Curie actions (grant agreements 742559 and 766025). We thank the Russian Science Foundation (Grant No. 20-62-46026) and Foundational Questions Institute Fund (FQXi) via Grant No. FQXi-IAF19-06 for supporting the work.

\end{document}